\documentclass[12pt,referee]{aa}

\usepackage{epsfig}
\usepackage{amssymb}
\usepackage{gensymb}
\usepackage{tabularx}
\usepackage{amsmath,amssymb}
\usepackage{amsfonts}
\usepackage{bbold}
\usepackage{graphicx}
\usepackage{color}

\usepackage[utf8]{inputenc}

\usepackage{color}
\definecolor{red}{rgb}{0.7,0,0}
\definecolor{blue}{rgb}{0,0,0.7}
\definecolor{purple}{rgb}{0.7,0,0.7}

\def\refe#1{{{#1}}}

\begin{document}
\title{Rossby Wave Instability and High-Frequency Quasi-Periodic Oscillations in accretion discs orbiting around black holes}

\author{Peggy Varniere \thanks{varniere@apc.univ-paris7.fr}\inst{1,2} \and Fabien Casse \inst{1,2} \and Frederic H. Vincent \inst{3}}

\titlerunning{GR RWI and HFQPOs observables around black holes}
\authorrunning{Varniere, Casse \& Vincent}
\institute{Laboratoire AstroParticule et Cosmologie, Universit\'e Paris Diderot, CNRS/IN2P3,  CEA/Irfu, Observatoire de Paris, Sorbonne Paris Cit\'e, 10, rue Alice Domon et L\'eonie Duquet, 75205 Paris Cedex 13, France
\and
Laboratoire AIM, CEA/IRFU-CNRS/INSU-Universit\'e
Paris Diderot, CEA DRF/IRFU/DAp, F-91191 Gif-sur-Yvette, France
\and
LESIA, Observatoire de Paris, Universit\'e PSL, Sorbonne Universit\'e, Universit\'e Paris Diderot, Sorbonne Paris Cit\'e, 5 place Jules Janssen, 92195 Meudon, France
}

\date{Received  / Accepted }

\abstract
{The rather elusive High-Frequency Quasi-Periodic Oscillations(HFQPO) observed in the X-ray lightcurve of black hole have been seen in a wide range of frequencies, even within one source. It is also notable to have been detected in \lq pairs\rq\ of HFQPOs with a close to integer ratio between the frequencies.}
{The aim of this paper is to investigate some of the possible observable that we could obtain from the having the Rossby Wave Instability (RWI) active in  the accretion disc surrounding the compact object.
}
{ Using the newly developed GR-AMRVAC code able to follow the evolution of the RWI in a full general relativistic framework, we explore how RWI can reproduce observed HFQPO frequencies ratios and if it is compatible with the observations. In order to model the emission coming from the disc we have linked  our general relativistic simulations  to the general relativistic ray-tracing GYOTO code and delivered synthetic observables that can be confronted to actual data from binary systems hosting HFQPOs.}{We have demonstrated in our study that some changes in the physical conditions prevailing in the part of the disc where RWI can be triggered leads to various dominant RWI modes whose ratio recovers frequency ratios observed in various X-ray binary systems. In addition we have also highlighted that when RWI is triggered near the last stable orbit of a spinning black hole, the amplitude of the X-ray modulation increases with the the spin of the black hole. Revisiting published data on X-ray binary systems, we show that this type of relationship actually exists in the five systems where an indirect measure of the spin of the black hole is available.}{}


\maketitle

\section{High-frequency quasi-periodic oscillations in X-ray binary systems}
\subsection{Introduction}

While fast variability from  black hole binaries is often detected, the most sought after observations are  those  of HFQPOs.
Those HFQPOs appear as narrow peak(s) in the X-ray power-density spectra (PDS) of black hole binaries and have been detected in eight different 
 black hole sources which are GRO J1655$-$40, GRS 1915+105, XTE J1550$-$564, H1743$-$322, 4U 1630$-$47, XTE J1650$-$500, 
 XTE J1859+226 IGR J17091$-$3624 \citep[e.g. ][and references therein]{Remillard06,altamirano12,belloni12}, ranging from 
{as low as 27~Hz in GRS 1915+105  \citep{belloni2001} } up to a  few hundred Hz \citep[a maximum of 450 Hz has been detected in GRO J1655$-$40][]
{Remillard06}.  HFQPOs are  particularly interesting as their frequencies typically lie in the frequency range of the Keplerian 
 frequency of the last stable orbit around the central  black hole. They can therefore be seen as a window 
 to the innermost region of accretion where strong gravity is expected to play an important role\footnote{Relativistic effects, and their influences on the properties of low frequency QPOs, are, for example, discussed within the context of the accretion-ejection instability in \citet{VTR12}.}. 
 This is one of the reasons 
 why HFQPOs {have stimulated much more interest than their low frequency counterparts (LFQPO)} even though  they are 
 much weaker and rarer.
 
 Another reason behind the interest in HFQPOs is that they sometimes exhibit an integer ratio  between detected peaks, such as 
 the classical 1:2  in GRS 1915+105 \citet{belloni_alt2013} but also a 2:3 ratio  in, for example,  GRO J1655$-$40 \citep{remillard02b} or 
 both the 2:3 and potentially 3:4 ratio in XTE J$1550$-$564$ \citep{miller01_1550,VR18}.  Those occurrences lead to a wide variety of models. Here we focus
 on one of them based on the  Rossby-Wave Instability (RWI).

        The  RWI, while  first introduced in galactic disc by \citet{Lovelace78}, was  rarely used because of the difficulty  
	 finding a physical setup fulfilling its criterion. 
 	It is only in the past decade that the RWI has been more widely used as physical settings where it could be triggered were discovered
	 from  planet-formation \citep{VT06,Meheut10,Lyra12, Lin12}  to the flares of Sgr A$^\star$ \citep{Tagger06,Vin14} and the 
	 fast variability of microquasars \citep{TV06,VTR11,VTR12,Vin13}.
	 That last case comes from the fact that the epicyclic frequency goes to zero at the last stable orbit, hence ensuring that the criterion for the 
	 RWI is fulfilled inside the inner region of the disc.	 

	This leads us to study in a full general relativistic framework the existence and evolution of the RWI  in the case of a Schwarzschild 
	\citep{CV17} and Kerr \citep{CV18} black hole. Here we take this a step further 
	by adding general relativistic ray-tracing to our simulation in order to create numerical observations that can be translated into observables such as the root mean square (hereafter rms) amplitude of the 
	modulation created by the instability, its quality factor and the presence of integer ratio peaks in the power-density spectrum (PDS).

\subsection{ {The Rossby Wave Instability as a model for HFQPOs}}	
\label{sec:RWI}
	
	 {Here we will review the salient  points  that makes the RWI a good candidate to explain the HFQPOs observed in microquasars. See  \citet{Lovelace14} 
	for a full review of the RWI in its different domain of application.}
	
        {{\tt $\bullet$}  Thanks to the vanishing of the epicyclical frequency at the last stable orbit, this hydrodynamical instability occur naturally in any disc getting close to 
      its last stable orbit \citep{TV06}.}
	
	  { {\tt $\bullet$} When fully developed, RWI may exhibit several unstable modes, each one of them  being characterised by a different toroidal integer numbers $m$. The strongest, dominant, mode of the RWI is not always the $m=1$ but higher order modes can be, simultaneously or not, present 
	giving to a natural explanation for the different integer ratios observed \citep{TV06}. This stems from the fact that the frequency induced by a RWI mode is directly proportional to the value of $m$.}

 	   { {\tt $\bullet$}  The RWI was shown to be able to occur simultaneously with one of the instabilities proposed to be at the origin of the LFQPOs 
	\citep{VTR11,VTR12}. This is important as HF and LFQPOs are often observed together. Not many HFQPO models have demonstrated their ability to exist 
	in a disc exhibiting LFQPOs.}

 	   { {\tt $\bullet$} The RWI is one of the few HFQPO model that has demonstrates its ability to modulate the X-ray flux up to the observed levels 
	\citep{Vin13}.}

         {In order to complete the next step in our exploration of the RWI characteristics and how they relate to the observed HFQPOs, we turn to 
	numerical simulations  to obtain more direct informations about observables such as the actual PDS peak distribution that rises from the RWI  and what it depends
	on, but also how the rms amplitude change with the spin of the black-hole.}

\section{ {{\tt NOVAs:} General relativistic simulations of the RWI around spinning black holes}}
 
         {\citet{CV18} performed the first general relativistic hydrodynamical simulation of the RWI in a Kerr metric context proving its existence in the disc around maximally  spinning black-holes.
        It also concluded that, while there were only limited general relativistic effects on the instability in itself,  the effect was mostly related to the time lapse and therefore hint
        at the necessity to ray-trace in a general relativistic context the emission back to the observer if one wants to look into more details at the impact of the spin. 
        For that reason we build the Numerical Observatory of Violent Accreting systems, {\tt NOVAs} \citep{VCH18XMM} which couple together the results from our}
        newly developed MPI\refe{\footnote{\refe{Message Passing Interface used for parallel computing.}}} code GR-AMRVAC \citep{CV18} with 
	the open-source general relativistic ray-tracing code GYOTO~\citep{Vin11} to create the lightcurves and then the PDS
	 associated with our simulations. 
	 Using those we are able to track the impact of the spin and the shape of the  {inner} region {, where the RWI is active,} on the flux modulation.

 \subsection{General relativistic  {hydrodynamical} simulations}
       {The GR-AMRVAC code solves general relativistic hydrodynamics equations including conservation of mass and  momentum. Spacetime geometry around a rotating, 
      uncharged black hole is fully determined by the Kerr metric \citep{Kerr63}. The general expression of any line element $ds$ in a (3+1) splitting in Boyer-Lindquist 
      coordinates is
      \begin{equation} 
ds^2= -\alpha^2(cdt)^2 + \gamma_{ij}(dx^i+\beta^icdt)(dx^j+\beta^jcdt)\nonumber
\end{equation}
 where $\alpha$ is the lapse function, $\beta^i$ is the shift vector and $\gamma_{ij}$ is the spatial metric tensor \cite[see e.g.][for definition of these quantities]{CV18}.
  In our notation greek letter stands for indices ranging over all coordinates while latin letter are restricted to spatial coordinates only.  We denote $\gamma$ as the determinant of the spatial metric tensor.   }
      
    {    Defining $\rho$ as the proper density of the 
      gas and ${\rm v^i}$ its contravariant eulerian velocity components (normalized to the speed of light $c$) one can express the aforementioned conservation laws as 
       \begin{eqnarray}
 \partial_t (\sqrt{\gamma}D) +\partial_j\left(\sqrt{\gamma}D\left(\alpha\text{v}^j-\beta^j\right)\right) = &0&\nonumber\\
 \partial_t (\sqrt{\gamma}S_i) + \partial_j\left(\sqrt{\gamma}\left[S_i(\alpha\text{v}^j-\beta^j)+\alpha P\delta_i^j\right]\right) = &&\nonumber\\
 \sqrt{\gamma}\left\{-(W^2\rho hc^2-P)\partial_i\alpha + \frac{\alpha}{2}\left(S^j\text{v}^k+P\gamma^{jk}\right)\partial_i\gamma_{jk} \right. && \nonumber\\
 \left. + S_j\partial_i\beta^j \right\}  && 
 \label{SetGRHD}
 \end{eqnarray} 
where $W=(1-{\rm v_iv^i})^{-1/2}$ is the Lorentz factor of the gas, $D=W\rho$ is the relativistic mass density, $W^2phc^2$ is the relativistic enthalpy of the fluid and $P$ is the thermal pressure.}  {One can easily define the relativistic momentum of the fluid as $S_i=W^2\rho hc^2{\rm v_i}$. The full energy conservation  includes heating and cooling mechanisms which would require considering physics beyond hydrodynamics such as high-energy kinetic theory and radiative transfer. Instead we choose to replace this equation by a much simpler power-law pressure prescription $P=C_o\rho^{\tilde{\gamma}}$ where $C_o$ and $\tilde{\gamma}$ are two positive constants. In order to close the system, we use the standard relativistic equation of state described by \citet{Mignone07}.} \\
        
        			  \begin{figure*}[htbp]
\centering
{\resizebox{9.cm}{!}{\includegraphics{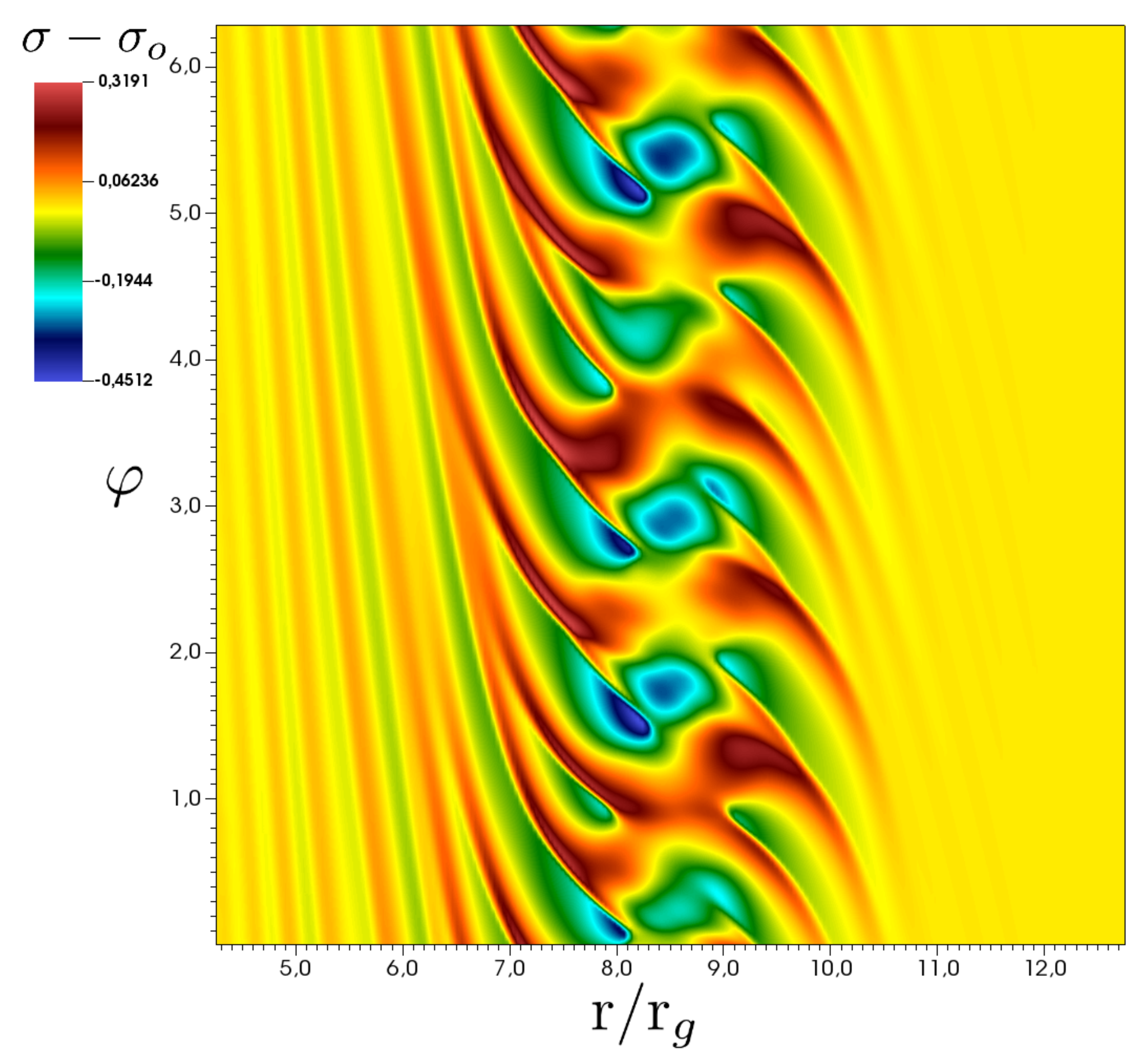}}}
{\resizebox{8.3cm}{!}{\includegraphics{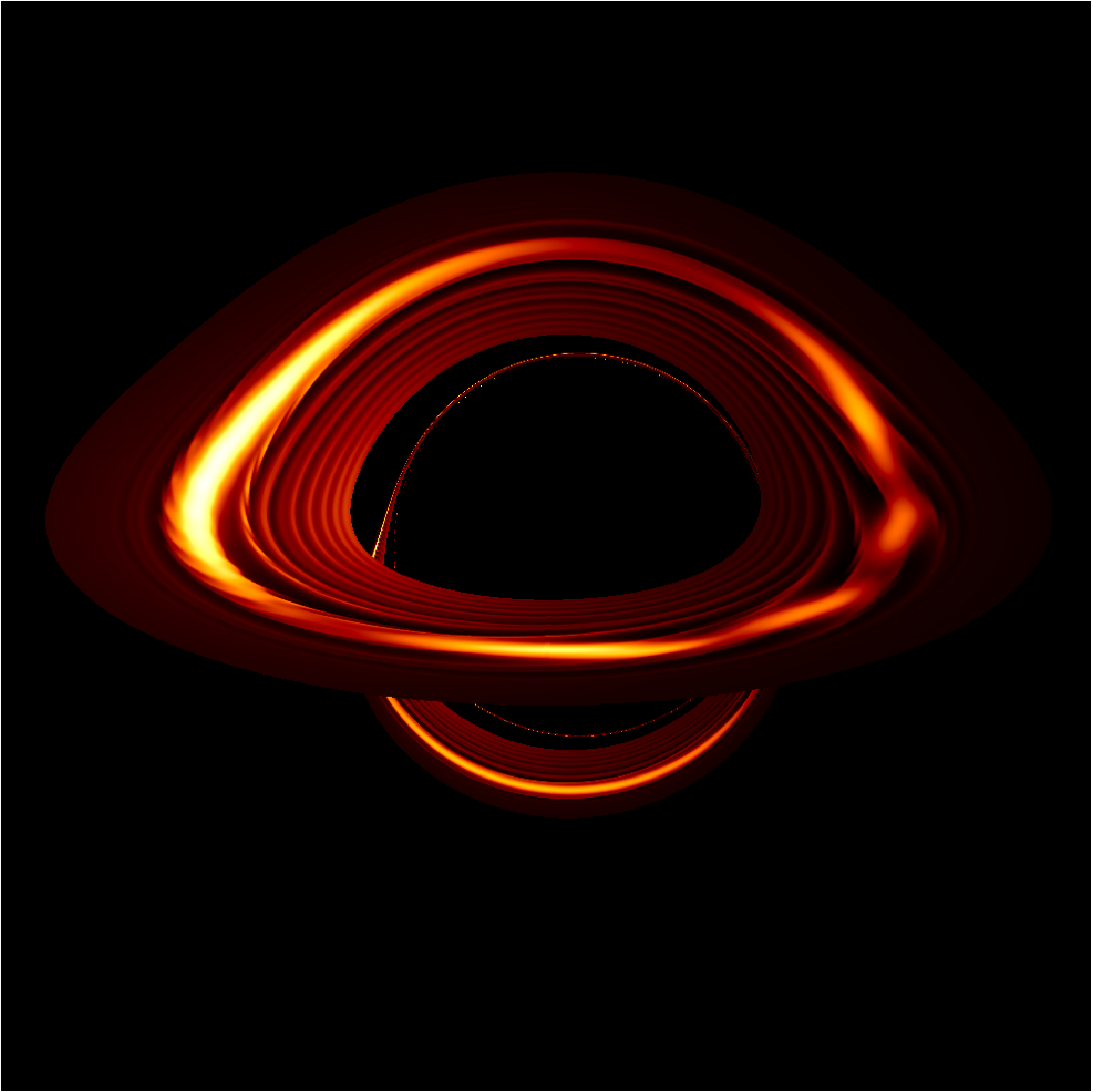}}}\\
\caption{\refe{{\bf Left:} A typical example of RWI triggered in an accretion disc at $r_b=8.5 r_g$ around a black hole whose spin parameter is set to $0.5$. Density perturbations are displayed exhibiting the occurrence of vortices located at $r_b$ from the black hole at a late stage of the simulation.  Identified vortices are all anti-cyclonic according to the rotation of the disc as anticipated when one considers the RWI. {\bf Right:} Synthetic image from the simulation displayed on the left panel. The GRHD simulation is performed by the GR-AMRVAC code while the image is processed using the GYOTO general relativistic ray-tracing code. This image considers photons (with energy of $2$ keV) reaching a remote observer standing far from the black hole.}}
\label{fig:Example_sim}
\end{figure*}

  { We use a similar setup as in \citet{CV18} for our simulations. These initial conditions are designed for vertically integrated accretion disc configurations. We adopt such approach 
 since accretion discs in X-ray binaries are likely to be very thin. We have already shown in \citet{CV17} that full 3D simulations of thin accretion discs prone to RWI lead to the same 
 results than vertically integrated simulations in the case of a non-spinning black hole. The vertical gravitational  force generated by Kerr black holes exhibits a similar behaviour  than 
 around Schwarzschild black holes when close to the equatorial plane. We can then assume that the relevance of vertically integrated simulations is still valid for simulations dealing
  with very thin discs orbiting around Kerr black holes. This assumption is also sustained by the fact that RWI vortices are essentially 2D structures when this instability is triggered 
  in a carefully crafted setup where no other instability is present \citep{Meheut12}.} \\
  
  	 {Following the criterion of the development of RWI, the unstable zone will then be the region where we have an extremum of the so-called vortensity (${\cal L} = \varepsilon^{\theta\rm i}_{\rm j}{\rm v}^j_{;i}/{\sigma(r)}$, $\varepsilon^{\i\rm j}_{\rm k}$ being a Levi-Civita tensor).}
    
  {  The initial conditions defining our simulations consider a surface density $\sigma(r)$ of the disc exhibiting a bump centered around the  so-called corotation radius $r_c$. }
  {The surface density of the disc is set in a similar fashion as in   \citet{CV18}, namely
\begin{equation}
 {\sigma(r)}= \displaystyle\left(\frac{r_o}{r}\right)^{0.7}\left(1+\varepsilon\exp\left\{-\left(\frac{(r-r_c)^2}{2\delta^2}\right)\right\}\right)
\label{Eq:Densdistri}
\end{equation} 
where $r_o$ corresponds to the last stable orbit radius for the fastest spinning black hole considered ($a=0.995$), $r_o=1.34\ r_g$. We ran a much smaller set of spins and positions, while changing the parameters of the unstable zone\footnote{ In  \citet{CV18} the unstable zone was kept unchanged to study only the impact of the spin. 
We refer to it to see the full impact of the spin and GR effects on the instability. Here we focus on observables.} $\varepsilon$ and $\delta$. With the profile chosen for those simulations we can succinctly parametrize the unstable zone by the amplitude of the density bump ($\varepsilon$) and the radial ($\delta$) extend of the fulfilled RWI criteria. } \\

 { The radial equilibrium of the disc is achieved when the centrifugal force balances both the gravity of the black hole as well as the     
 thermal pressure of the gas. This leads to the prescription of the rotational velocity as 
   \begin{equation}
 {\rm v}^{\varphi}=\displaystyle\frac{-\partial_r\beta^\varphi+\sqrt{(\partial_r\beta^{\varphi})^2+2\alpha\partial_r\gamma_{\varphi\varphi}(\partial_r\alpha+\alpha\frac{\partial_rP}{\rho hW^2})}}{\alpha\gamma^{\varphi\varphi}\partial_r\gamma_{\varphi\varphi}} 
 \label{Eq:shear}
 \end{equation}
 The radial pressure profile $P$ is set following the aforementioned power-law prescription where the parameters $C_o=1.8\times 10^{-4}$ and $\tilde{\gamma}=3$ as in \citet{CV17,CV18}. The values have been chosen in order to be consistent with equatorial conditions consistent with a thin disc whose aspect ratio is $H/r\sim 4\times 10^{-2}$. Let us note that in our vertically integrated disc context, the disc scale height $H$  is evaluated using the assumption of a vertical hydrostatic equilibrium  where the vertical pressure gradient is expected to balance to vertical gravity. }
\\
 
 {The instability is then triggered by injecting very small amplitude random velocity fluctuations near the corotation radius.}
	In all the simulations, the disc goes from its last stable orbit to about 3 times that, while we manually centered the unstable zone  
	by choosing the aforementioned  density profile. 
	The grid resolution is $196\times 600$ for each simulation while the spatial domain ranges from $r_{LSO}$ to $3r_{LSO}$ and $\phi \in [0,2\pi]$.  
	The radial boundaries are continuous while azimuthal boundaries are periodic. In order to time advance, we used a Harten, Lax and van Leer (HLL) 
	solver linked to a Koren slope limiter. A typical simulation requires approximately several tens of thousands time steps and 
	lasts a few of hundreds cpu-hours.\\ 
 
 	\refe{The left figure of Fig.\ref{fig:Example_sim} shows the density profile for the case of a=0.5 with the RWI triggered at $r_c=8.5r_g$. As we are using a subset of our simulations
	presented in  \citet{CV18}, we refer the reader interested in a more in-depth study of the RWI behavior to that paper. Here we will focus more on the possible observational
	consequences of the GR effects.}

 \subsection{General relativistic ray-tracing to the observer}

	For all ray-tracing computations in this article, we use the open-source\footnote{Freely available at \url{http://gyoto.obspm.fr}}
	GYOTO code.  Photons are traced  {backwards in time} by integrating the geodesic equation using a Runge-Kutta-Fehlberg adaptive-step integrator at
	order 7/8 (meaning that the method is 8th order, with an error estimation at 7th order). 
	From such maps of specific intensity, the light curve (flux as a function of time) is derived by summing all pixels weighted
  	by the element of solid angle, which is subtended by each pixel. GYOTO computations consider the very same Kerr metric than the hydrodynamical 
	simulations so that all general relativistic effects upon photons are taken into account in our study, namely gravitational frequency shift and time lapse.\\
	
	 { A backwards-integrated photon is traced until the accretion disc is reached.
	The local temperature is derived from the 2D grid of the GRHD simulation
	by considering the relation
	$T(\rho) = T_{\rm inner}  (\sigma/\sigma_{inner})^{\gamma_o-1}$
	where $\sigma$ is the local density interpolated from the grid, $\sigma_{\rm inner}$ is the
	density at the inner radius of the disc, $T_{\rm inner}$ is the temperature at
	the inner radius, fixed at $T_{\rm inner}=10^{7} K$, and $\gamma_o$ is the adiabatic index,
	fixed at $\gamma_o=5/3$. The emitted intensity is then simply the Planck function
	evaluated at this local temperature. The local velocity of the emitting gas is provided by the GR hydrodynamical simulations, 
	which allows to compute the redshift factor and translate the
	emitted intensity to the intensity as observed by the distant static observer.} \\

	As we are only interested in studying the impact of the disc, or more precisely the parameter of the unstable zone, we look at a black hole with an inclination with 
	respect to the observer of $70^\circ$. As more than half of the HFQPO sources have a high inclination it is not a strong restriction. 
	Changing the inclination has for consequences
	changing the strength ratio between peaks but not their distribution.  
	 \refe{On the right of Fig.\ref{fig:Example_sim} we show a typical disk image for an inclination of 70$^\circ$ for the case of a=.5 and the RWI occurring at $r_c=8.5r_g$.}

\section{ {What causes the PDS mode selection}}

	 {We saw in sect.\ref{sec:RWI}.  that one of the interesting feature of the RWI to explain HFQPOs is the fact that the strongest dominant mode is not always the 
	$m=1$ and that multiple higher modes can occur \citep{TV06} but this was studied in semi-analytical form only. Here we use {\tt NOVAs} to explore numerically 
	the different PDSs we can get from the RWI while taking into account all physical effects induced by the gravity of the black hole.  }

	 {This is of particular interest as
	several integer ratios between the frequency of the observed HFQPOs have been detected,  sometimes in the same objects\footnote{It is important
	to note that, most of the time, those are not detected in individual observations but while accumulating several observation together with different criteria. See
	for example \cite{miller01_1550,remillard02b,VR18} for the occurrence of the 3:2 and 3:4 ratio in XTE J1550-564.}. It is therefore interesting to study the changes
	required for the RWI to show such different ratios in its PDS and see if those are compatible with observations.}

\subsection{Synthetic Observations  {of the RWI exhibiting 3:2 and 3:4 ratio}}
\label{sec:mode}

	 {From analytical and semi-analytical work \citep[see for example][]{Lovelace14} we know that the local conditions where the RWI grows}
	 are what trigger the growth of a mode over another, and in turn it will directly impact the PDS.
	In the following section we will focus on only one spin ($a=0.995$) and position ($r_c=2.7 r_g$)  {where the RWI is triggered} 
	but look at different  {local conditions (see Eq.\ref{Eq:Densdistri} for the parametrization), meaning the}  (unstable) zone  {where the RWI is active will be different}.
	 {We can then} see how the  {local} condition
	affect the PDS especially focusing on which modes are selected and how they relate to one another when present. 
	We saw in previous studies of the RWI \citep{TV06,Vin13,CV17,CV18} that every simulation triggers a lot of modes, with up to $m=19$ being detectable in Fourier space,
	though lower modes tend to be a lot stronger, especially once the instability reaches saturation. Because of the timescale near the last stable orbit of a ten solar mass black-hole, 
	we will be mostly observing the saturated state of the instability in a standard observation. We have then first performed several simulations having the same aforementioned corotation radius 
	$r_c=2.7 r_g$ but with various $\varepsilon$ and $\delta$ parameters leading to different geometry of the unstable zone. 
	\refe{A typical GRHD simulation example and the related ray-tracing synthetic observation can be found in Fig.\ref{fig:Example_sim}.}  Once a simulation has reached the  saturation stage, namely when the amplitude of the fluctuations remains approximately constant, we stop the simulation and proceed to use  {GYOTO} in order to compute the light curve of the simulation as perceived  by a remote observer. We stress here that the combination of the { GR-AMRVAC} code and the { GYOTO}  code encompasses all general relativistic effects at work upon both the disc and its radiative emission, here providing  reliable synthetic observations.   
\newline

	  {Using the combination of { GR-AMRVAC} and { GYOTO} we obtain the lightcurve and then the Power Density Spectrum (PDS) where the different
	 peaks of HFQPOs can be detected.}
\begin{figure}[ht]
\centering
{\resizebox{8.5cm}{!}{\includegraphics{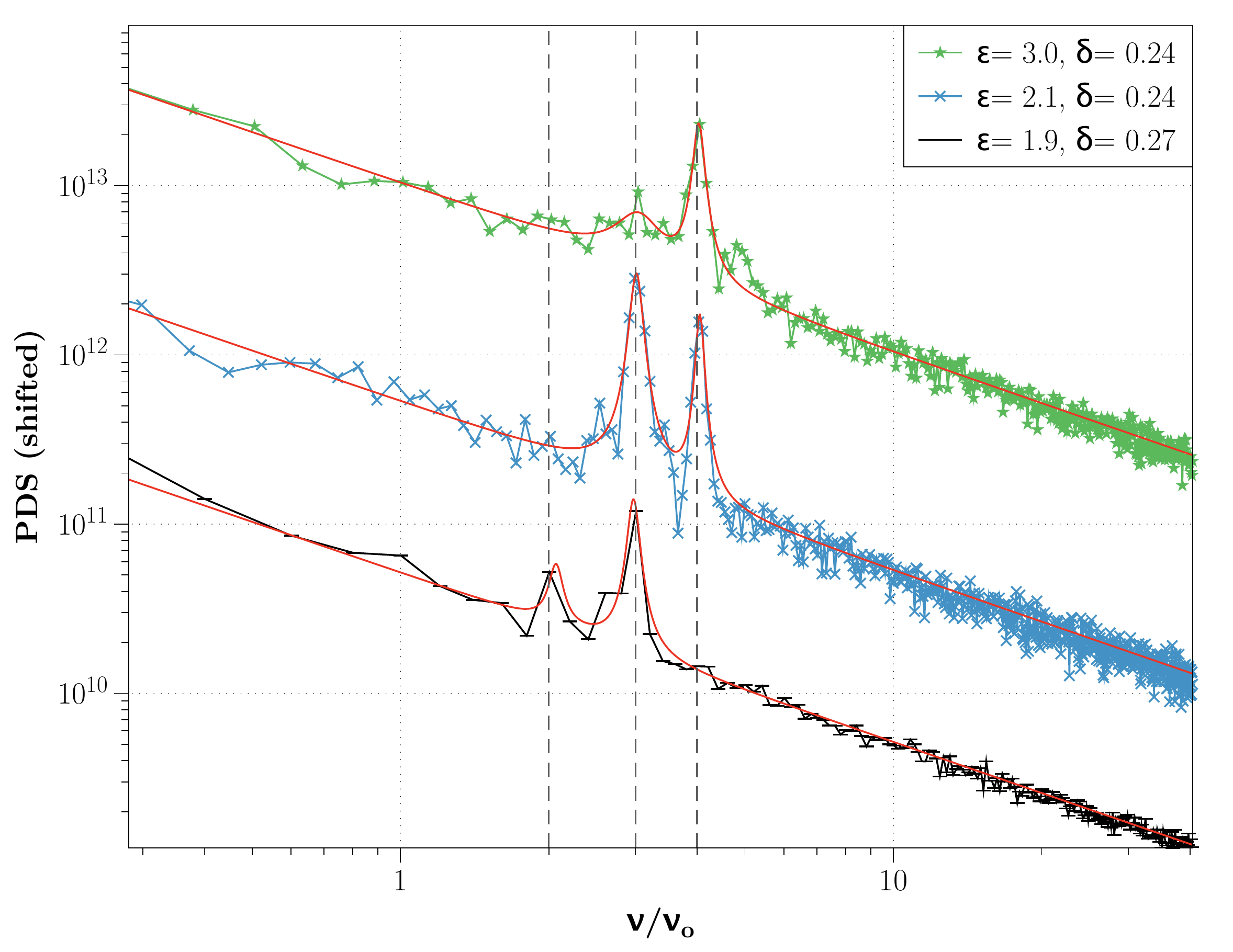}}}\\
\caption{Three examples of resulting PDS for the same spin ($a=0.995$) and position  ($r_c=1.35 r_S$) but with different  {parameters of the system.} For visibility the PDS are 
vertically shifted.} 
\label{fig:PDS_diffbump}
\end{figure}
	 When computing those PDS from the simulations we often get only one dominant peak or with a small, close to the detection level, secondary peak, as we can see 
	 on the upper curve of Fig.\ref{fig:PDS_diffbump}.  In a few cases\footnote{ {We did not run enough simulations to statically determine the rarity of one case compare
	 to another.}}  {we see that} multiple peaks are well above the detection level.
	 Fig.\ref{fig:PDS_diffbump}. shows examples  {of the most commonly observed integer ratio in black-hole binaries, namely the} 2:3 and 3:4 ratio, both of those having been observed 
	  respectively in GRO J1655-40 \citep{S01a} and in XTE J$1550$-$564$ \citep{VR18}.

	 It is interesting to note that  {for} XTE J$1550$-$564$ a case of 2:3 has also been detected by \citep{remillard02b} by adding together observations having similar
	 low-frequency QPO types\footnote{How using the LFQPO characteristics to uncover the HFQPO's behavior relates to the instability presented here has been shown
	 in more detail in \cite{VTR12}.} making it the first source with multiple integer ratios being detected.
 {Not only Fig.\ref{fig:PDS_diffbump}. proves the ability of the RWI to create a detectable modulation of the flux, but it also demonstrate that several integer ratio
	of the peak distribution could be observed in the PDS of the same object, depending on the local conditions in the inner region where the RWI is active.} 
	
	 {As we have not performed simulations for a wide range of parameters, we cannot be sure of the minimal level of change require in the system to get
	different peak ratios. Nevertheless,}
	 in the framework of the  {three} RWI simulation presented here, such differences in the  {integer ratio of the} peak distribution  {is associated with a relatively large}
	 change of the parameters of the unstable zone	 of about $30$\%.   {Meaning that a measured change of  $30\%$ in the parameters of the inner region is enough to trigger a change 
	 from 2:3 to 3:4 ratios in HFQPOs.}
	
\subsection{How much change in the observations?}
   
   	 {The physical variations in the inner region of the system required by the RWI in order to be able to explain a change in peak ratio is strong enough to not be straightforward.
	For that reason, we decided to look at the observations of} the source XTE J$1550$-$564$ which has, to date, the highest
	number of different detected HFQPOs along with several distinct integer ratio detections. Taking into account the 1998-1999 and 2000 outbursts, we have about thirty 
	observations with an identified HFQPO which is enough to see if  {the needed} variation of about $30$\%  {of the parameters of the inner region of the disc is compatible
	with the data.}

\begin{figure}[]
\centering
{\resizebox{8.5cm}{!}{\includegraphics{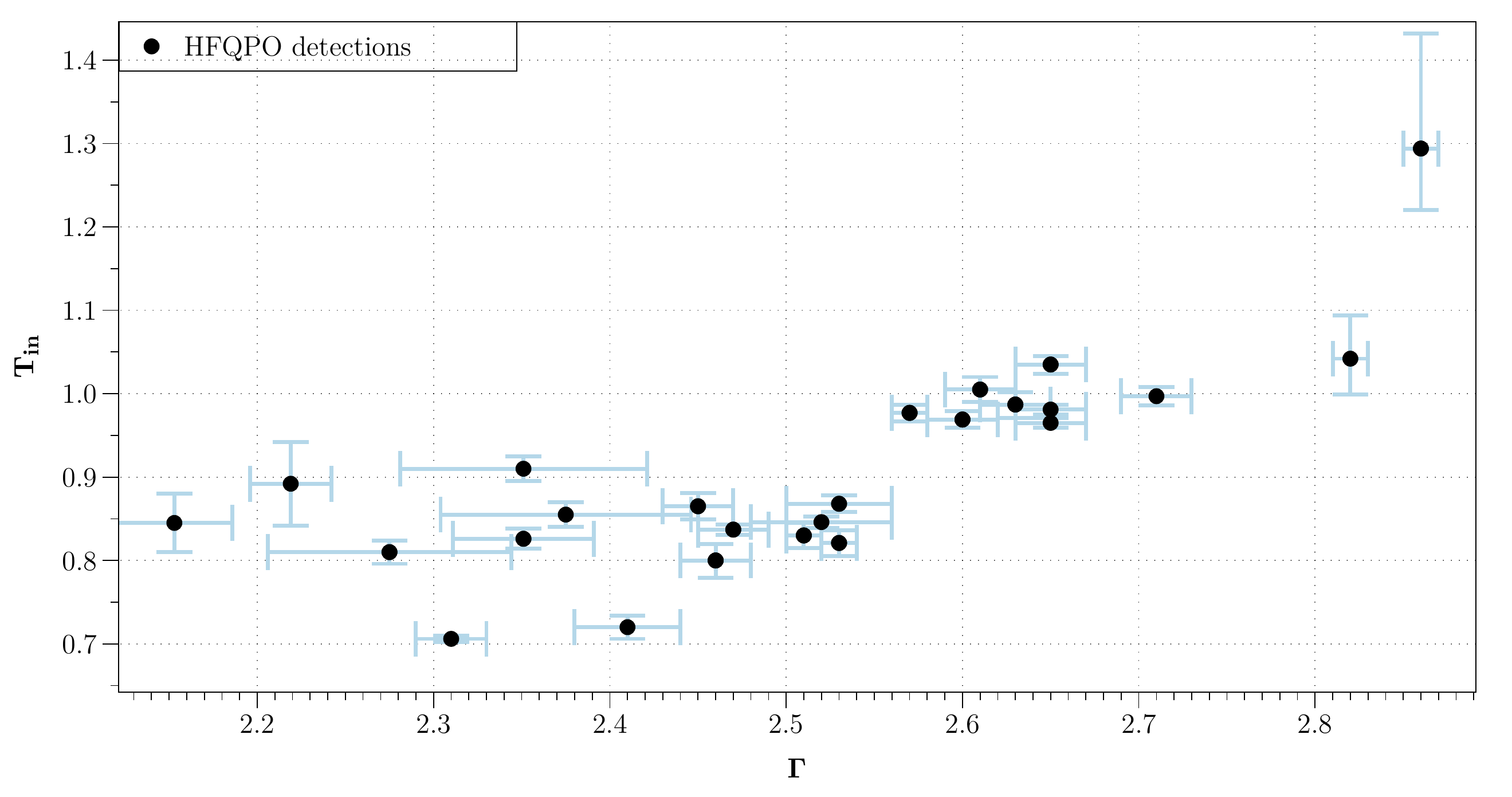}}}\\
\caption{the photon index $\Gamma$ versus the temperature at the inner edge of the disc for all the observations of XTE J$1550$-$564$ where a HFQPO was detected. We see that the 
temperature at the inner edge of the disc varies by about a factor of two while the photon index varies by close to $30$\%. Data from \citet{Sobczak00,remillard02b}. 
} 
\label{fig:obs}
\end{figure}

	 {Using RXTE, we rarely have}, other than the QPO itself, a direct link to the parameters of the unstable zone in the disc\footnote{ {See \cite{VMR16} 
	for more details on how we can get access to these parameters.}}.  Instead we focus on the spectral parameters associated with each
	observations, indeed the RWI is expected to arise near the last stable orbit and therefore impact the temperature at the inner of the disc as well as the comptonized component.
	  {Using data from \citet{Sobczak00,remillard02b},}  we plot in Fig.\ref{fig:obs}. the photon index $\Gamma$ versus the temperature at the inner edge of the disc for all the observations of XTE J$1550$-$564$ where a HFQPO was detected. We see that the temperature at the inner edge of the disc varies by about a factor of two while the photon index varies by close to $30$\%.
\newline

		While those changes in the spectral parameters do not imply a similar change in the  {inner region where the RWI will develops, it} will have to \lq respond\rq\  to those
		changes in the  {overall} disc and corona. Faced with a change in the temperature at the inner edge of the disc of about $100$\%, it is conceivable to have the unstable
		region, which is not very far from the inner edge of the disc, change by the factor of $30$\% needed for the RWI to explain a wide variety of modes. 
		As a result one can safely assume that a variation of typically  $30\%$ in the parameters of the inner disc cannot be ruled out by actual data from XTE J$1550$-$564$. 
		At the same time it also explains why multiple ratio in one system
		are so rare. Under such circumstances RWI naturally provides  an explanation for the HFQPOs frequency ratio shift within the same astrophysical system.

\section{Impact of the spin  {on the RWI saturation level}}	
	 
	Because of their high frequency, HFQPOs are often thought to originate from the inner region of a disc close to its last stable orbit, sometimes even requiring
	a non zero spin  {as it was argued by \citet{S01a} for the case of GRO J$1655$-$40$}, hence there is a link between the frequency of the HFQPO and the spin
	of the black-hole, but this link is hard to pinpoint observationally. Indeed, while HFQPOs sometimes appear in pairs of integer related frequencies making it easier 
	to identify the fundamental frequency, they most often do not and we do not know how the observed peak  relates to the fundamental frequency.
	
	Here we are exploring the possibility of another observable which can be impacted more directly by the spin of the black hole around which the RWI develops {,
	namely the maximum rms amplitude a HFQPO can attain}.\\
	
\subsection{ {Saturation of the RWI as function of spin}}	
\begin{figure*}[htbp]
\centering
\begin{tabular}{cc}
{\resizebox{10cm}{!}{\includegraphics{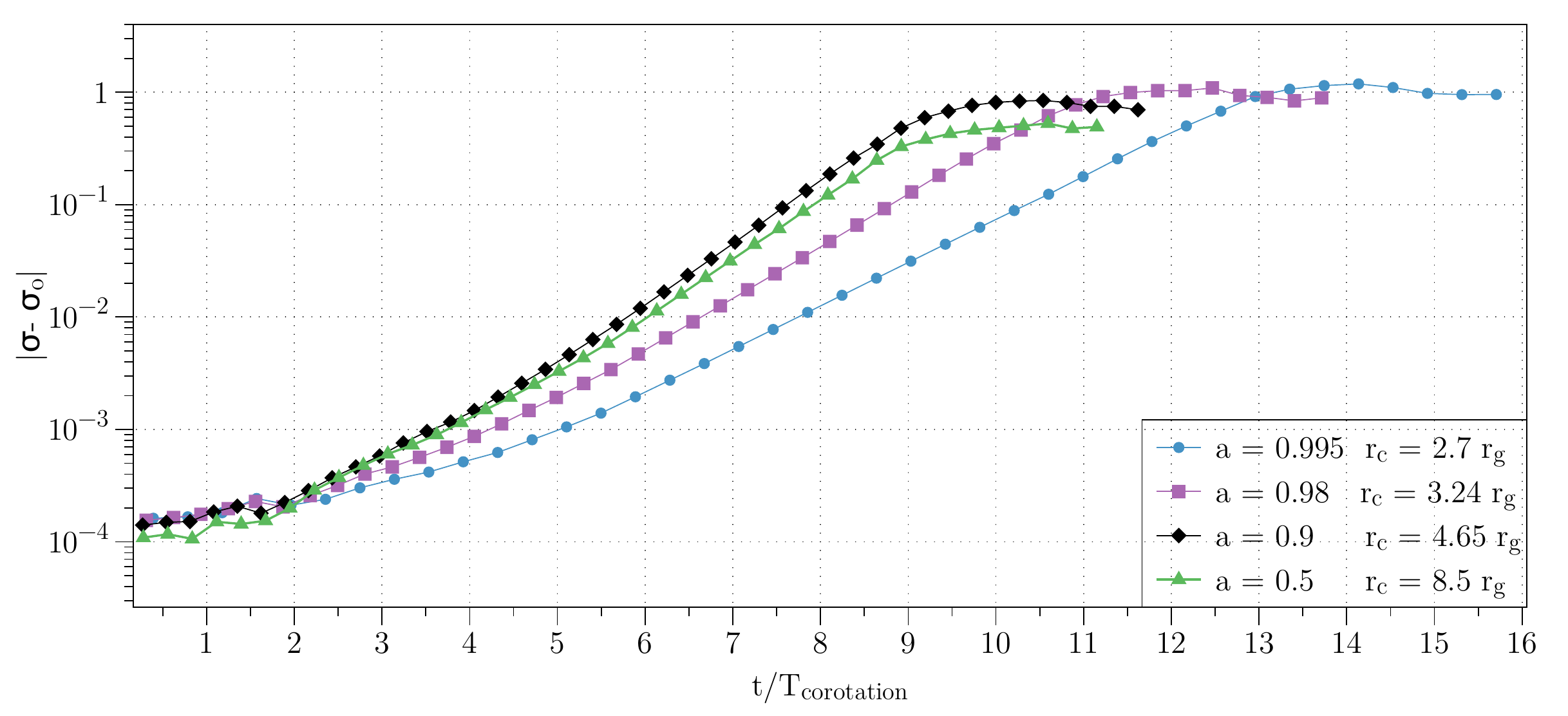}}} &
{\resizebox{7.3cm}{!}{\includegraphics{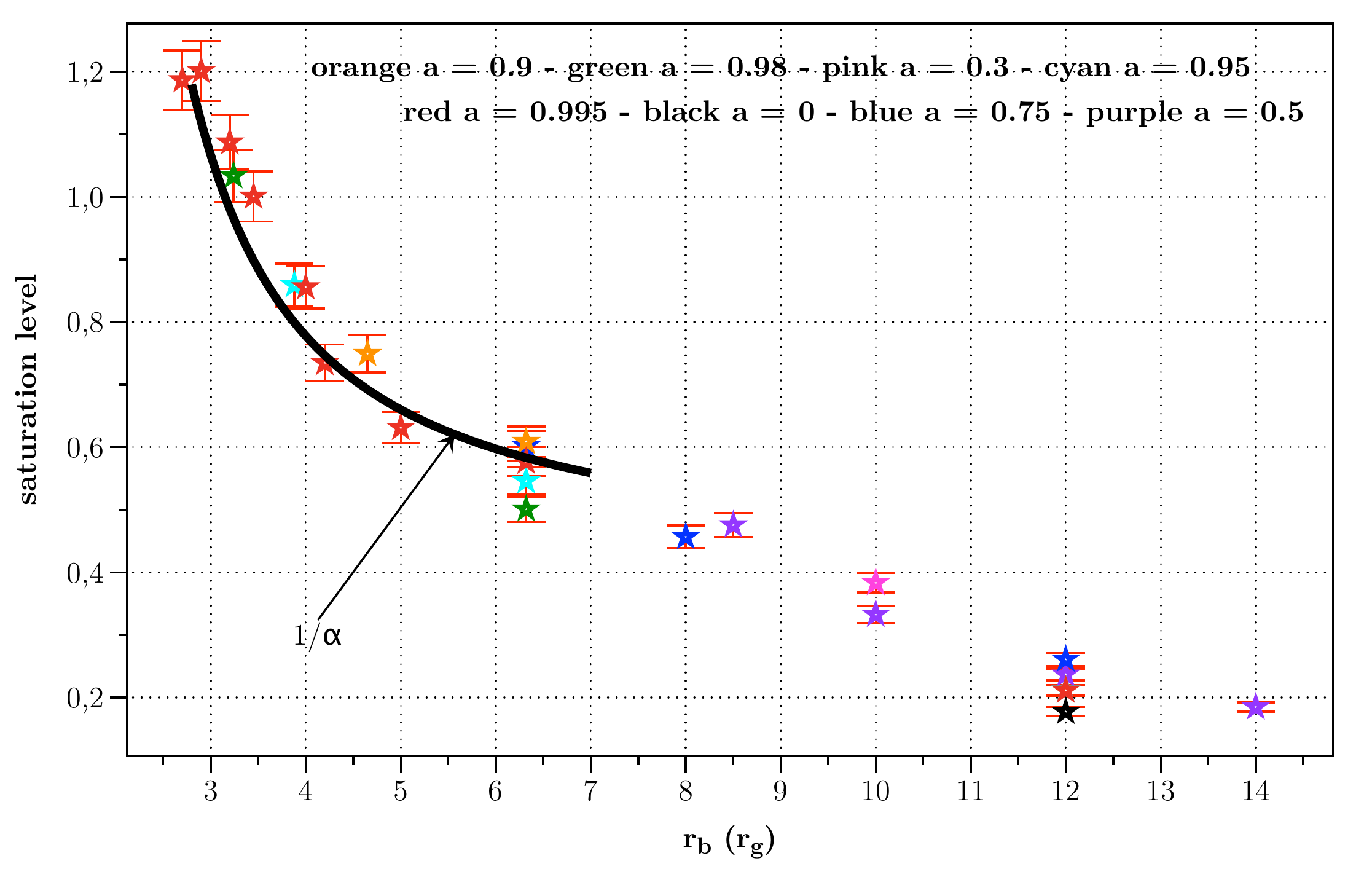}}}
\end{tabular}
\caption{{\bf Left:} Evolution of the maximum of $|\sigma-\sigma_o|$ as function of time normalized by the corotation time for four different spins, showing
how an increased spin causes a higher saturation (it is a logscale in y and the saturation has more than a factor of two differences.). The cases represented
here are, from right to left, blue circle $a=0.995$ at $r_c=2.7r_g$, purple square $a=0.98$ and $r_c=3.24 r_g$, black diamond $a=0.9$ with $r_c=4.65 r_g$ and green triangle
$a=0.5$ at $r_c=8.5r_g$. {\bf Right: } Saturation levels reached by RWI in our simulation sample with corotation radius and spin parameter ranging from $r_c=14 r_g$ to $r_c=2.7 r_g$ and from $a=0$ to $a=0.995$. All simulations have the same set of parameters regarding the unstable zone, namely same $\varepsilon$ and $\delta$. 
It clearly appears that the saturation level increases as the corotation radius is decreasing meaning that higher spin will lead to the RWI occurring nearer to the black-hole, which then lead to enhanced saturation level of the instability.
} 
\label{fig:rmsS}
\end{figure*}		
	 One of the major conclusions of \citet{CV18}  is that the saturation level of the RWI depends on the position where
	 the instability occurs but not the spin of the black hole, meaning that for the same corotation radius (i.e. same observed frequency), the spin of the black hole has only a marginal effect
	 on the maximum saturation the instability can reach. In this section  we will be using simulations that appeared in \citet{CV18}   and compare the saturation levels of different  spin/position 
	 couples,  when the  RWI develops in the inner region of the disc associated with that spin.

	 In the left panel of Fig.\ref{fig:rmsS} we show the evolution, and saturation, of {the maximum} of the surface density variation under the influence of the RWI
	 (computed by $|\sigma-\sigma_o|$, $\sigma_o$ being the equilibrium value of the density), 
	 as function of the time normalized by the corotation time for mild to extreme spin. 
	 We see that, as the spin increases the saturation level of the instability increases, as well as the time to reach saturation. While the change in time to reach saturation
	 might seem important (reach about $40\%$) when taking into account the actual timescale at the corotation, it is an extremely small difference, below what we can detect
	 at the moment. What is more interesting is the net increase in the saturation level (the plot is in log to see the behaviour from the start of the simulation) by more than 
	 a factor of two between mild to extreme spin. The right panel of Fig.\ref{fig:rmsS} summarises the saturation level obtained from all simulations performed with the same geometry of the 
	 unstable region, namely with the exact same value of $\varepsilon$ and $\delta$. On this figure one can easily see that the saturation level of the RWI is increasing as the corotation radius 
	 decreases. For $r_c > 4 r_g$ the increases follows the variation of the local density in the disc while for the closest corotation radii ($r_c < 4 r_g$) the increase is majored by general 
	 relativistic  effects. Having inner part of an accretion disc so close to the black hole is only possible for high value of the spin 
        so a prediction of the RWI model is that, for a similar setup, a higher spin, {\em i.e.} a smaller inner radius assuming the disk reach its last stable orbit, will mean a higher saturation level.
        
	This is not something we can straightforwardly compare with observations.  
	Indeed, in Fig.\ref{fig:rmsS}. the physical conditions in the  {system, or more precisely in the region where the RWI is active},
	are similar \footnote{ You can see how close are the disc parameters in see Fig.1. of \cite{CV18}.}
	while we have no  {simple} way to constrain how different, or not, are the  {inner disc} 
	conditions between two observations of two different objects. 

\subsection{ {rms amplitude of HFQPOs as function of spin}}	
	
	 {In order to assess a potential relationship between rms amplitude and the spin of the black hole, we can have a look at the overall behavior of the rms of HFQPOs in 
	all the objects we have known spins for.} By taking into account the overall 
	 {envelope} shape of the rms versus spin	we get the  {evolution of the} \lq maximum\rq\   (to this day) cases for all the objects. 
	If we assume that this maximum  {is reached, for} each object,  {in} similar  {conditions} in term of disc properties, then 
	using Fig.\ref{fig:rmsS}.,  {the RWI predict}  the maximum achievable rms to increase with the spin of the black hole. 
	 {This is something we can check with the limited sample of spins estimates we have for HFQPO system. }

\subsubsection{ {Spin determination in observations}}

	 {As we are using only published values and did not redo any of the data reductions, there exist only a limited sample of sources having both a spin measurements, 
	by any methods, and an observed HFQPO with published  rms value.}
	 {Here we will be using spin estimate from one (or both) of the two main methods, namely the continuum fitting (CF) and the iron line fitting (FeK). 
	Both methods have been shown to have limitations, indeed, in order to get the spin from either fit we need to have a good grasp on the disc fitting model \citep{KD11}.
	Keeping in mind, the known limitations, we can still use the limited sample of published estimates to compare with this prediction of the RWI  
       to see if there is an overall agreement or contradiction.} \\

	 {Our sample is composed of }
	4U1630-472 (red) with data from \citet{King14,KleinWolt04} ,
	H1743 -322 (blue) with data from \citet{Steiner12,Homan03},
	XTE J1550 -564 (green) with data from \citet{Steiner2011,remillard02b}, 
	 GRO J1655-40 (purple) with data from \citet{Shafee06,Reis09,S01a} and 
	XTE J1650-500 (black) with data from \citet{Miller02,Minutti04, Walton12,Reis13,Homan03}.

\begin{table*}[h]
\begin{center}
\begin{tabular}{c||c|c|c||}
Source & CF & FeK & used \\
\hline
4U1630-472       &    & $>$ 0.95     & $>$ 0.95     \\
H1743-322        & 0.2 $\pm$ 0.3  & & 0.2 $\pm$ 0.3\\
XTE J1550-564 &  0.34 $\pm$ 0.28  & 0.55 $\pm$ 0.1 & 0.36 $\pm$ 0.29\\
GRO J1655-40   &  0.7 $\pm$ 0.1  &  > 0.9  & 0.8 $\pm$ 0.2  \\
XTE J1650-500  &   & 0.84-0.98 /  $\sim$ 0.998 & 0.84-0.998  \\
                          &   &  $\geq$ 0.93 / $0.977_{- 0.07}^{+0.06}$ &   
\end{tabular}
\end{center}
\caption{ {Values of the spins for the five HFQPO source that have a spin estimate either from \refe{the} continuum fitting \refe{and the infered position of the last stable orbit} (CF), 
the iron line \refe{and reflection}  fitting  model (FeK) or both. 
The last column are the range displayed on Fig.\ref{fig:rmsO}., when several value exist, we choose a range that encompass all values.}}
\label{tab:spin}
\end{table*}

\begin{figure*}[t]
\centering 
{\resizebox{13.95cm}{!}{\includegraphics{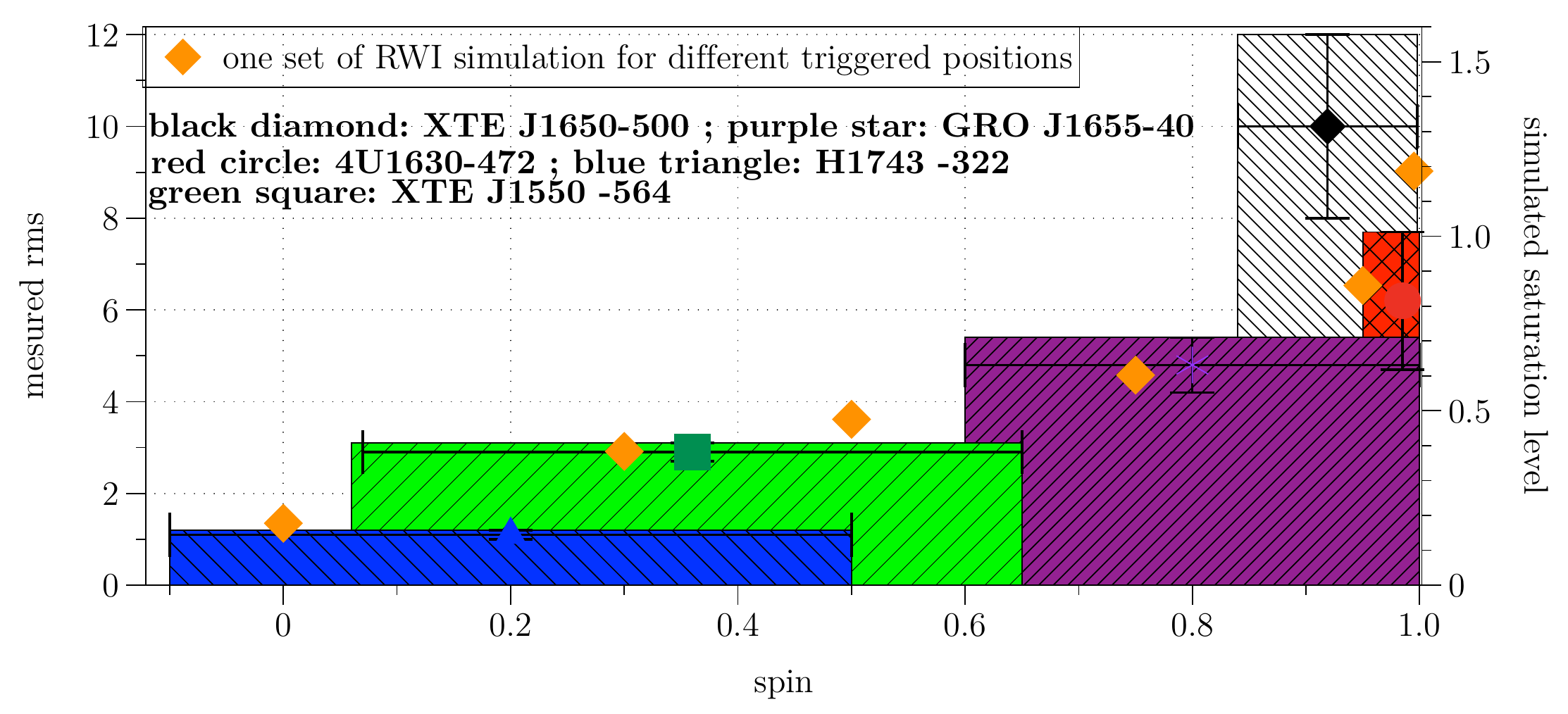}}}\\
\caption{Evolution of the rms amplitude of published HFQPOs as function of spin, showing  {a behaviour compatible with}  higher spin source tending to have higher maximum rms. See the text for references.
 {For comparison we added the value of the simulation setup from Fig.\ref{fig:rmsS}. as orange diamond. We see that, within the error bar, the observed data seems to follow
a similar evolution as function of spin as the one predicted by the RWI.}}  
\label{fig:rmsO}
\end{figure*}	

	As we can see from  table \ref{tab:spin}, in our sample we have two objects with only one determination while two objects have spin estimates from both
	 methods and one object has been measured several times with the same method but by different groups on different observations. 
	 We get a good agreement of both methods for XTE J1550-564 and we see that three of the four iron line fitting spin estimation of XTE J1650-500  are 
	 coherent with one another, while the last one is  a little outside two of the estimate (but in agreement with the last one). 
	 The only real disagreement we have between the two methods is for GRO J1655-40 where \refe{the variability (HFQPOs) and the inferred position of the last stable 
	 orbit} method provides a spin of  0.7 $\pm$ 0.1 \refe{\citep{S01a}} while the 
	 \refe{use of the disk reflection signatures} requires the spin to be  higher than 0.9 \refe{\citep{Reis09}}.

	 Because our sample is so small with only 2 objects having spin agreement with both methods  we need to take into account all of the 
	 known objects having a spin assessment with one of the methods. Using all the data from \citet{McC14} and \citet{R14} we obtain a slightly bigger sample of seven objects with
	 spin estimates from at least one of the two methods.
	  Of those, four have coherent spin estimates  from various studies using the same method with different observations. One of the systems showing a spin estimate discrepancy, namely 
	  4U 1543-47, was studied by \citet{MM14}  using both technics simultaneously.
	  They found that the difference between the two methods came mostly from the hypothesis used, for example the disc inclination coming from the iron line fitting
	  and the one used in the  continuum fitting were not the same. \citet{MM14} showed that using the same inclination led them to obtain coherent spin measurement with both methods. The value
	  of the spin they found with this method was consistent with both other values within errors.  So we now have five out of seven systems with trustful spin values, which give
	  some credibility to the method. 

	  {It is worth considering that GRO J1655-40, the case that mismatch in our sample, is thought to perhaps have spin misalignment. As shown in  \citet{MM14} 
	 this could lead to the difference in the spin estimation and a more thorough estimate of the spin, taking onto account both methods, is needed here.} 

	   {As a result we decided to use a large range of spin value by expanding  the error bars in order to encompass all the available spin estimates.
	  Those are shown in the last column of table~\ref{tab:spin}.}

\subsubsection{ {Evolution of the maximum rms as function of spin}}

	With the rms published in the aforementioned publications and the spin estimate from table \ref{tab:spin}, Fig.\ref{fig:rmsO}. shows,  
	how the maximum rms evolves as function of the spin. We choose
	to represent the distribution of rms versus spin as column expressing visually the error bars with color boxes.

	This is a sparse sampling  with large error bars, but it shows agreement with the predicted behaviour of a RWI-based modulation, {\em i.e} 
	the maximum rms amplitude will increase as function of spin.
	For reference, we also added to the plot with orange diamond symbols the results from the set of RWI simulation shown in Fig.\ref{fig:rmsS}. {\em Those are only there to show the trend
	but do not represent a \lq fit\rq\  of the data as we do not know enough the conditions of the system at the time of the observation to run the adapted
	simulations.}

      For the moment we have only very few sources with known spin and few sources with known HFQPOs and the intersection is the five sources presented here. 
       It would be interesting to pursue this study by checking all the sources with known spins for HFQPOs and also focus the measure of spin on the sources known
       to harbour HFQPOs. This could almost double the number of points (we have eight known sources with HFQPOs detected as of now) which would lead more
       credence to the behaviour presented here.

\section{Conclusions}

	Thanks to  {NOVAs,} our newly developed GR fluid code and the associated general relativistic ray-tracing, we have created  synthetic observations of a Kerr black-hole system with 
	a RWI developing in the disc.
	This allowed us, for the first time, to study more directly observables from the RWI. Here we presented two predictions arising from associating the RWI with the
	cause for HFQPOs.
	
	The first prediction, which is now observed, concerned the ability, for the same system, to display several integer ratio frequencies in its power distribution spectra at different times.
	As this became observed, we turned to explore what is required in order to achieve this change. We showed that the RWI was able to reproduce the
	PDS with different dominant modes and integer ratios between them  {if the inner region of the system changed by about} 30\%.
	Using published data, we showed that such changes is compatible with observations. Using more precise observations, such as NICER data, 
	will help constraining the conditions required to detect certain frequencies/ratios.
	Indeed, more simulations, especially linked with higher resolution observations, are needed in order to pinpoint what is the main component behind 
	the mode selection of HFQPOs. 
	
	The second prediction pertains to the link between the maximally attainable rms and the spin of the black hole. Indeed, under the same setup, the RWI saturation
	was seen to increase with the spin of the black-hole. 
	This is more tricky as spin estimation is not something the entire community agrees on. Nevertheless, we could not pass up the opportunity to use the limited sample of 
	published estimates to compare with this prediction of the RWI.  Using published data we find that astrophysical systems in our sample exhibit maximal rms levels that are compatible 
	with the RWI prediction, namely that maximal rms level increases with the spin of the black hole.  We aim to further this by looking for more HFQPOs in sources with known spins,
	especially in the new NICER data.

\begin{acknowledgements}
We thanks the anonymous referee for his comments on how to improve this paper.
PV Thanks J.Steiner for his advice concerning the spin determinations in black-hole binaries.
We acknowledge the financial support of the UnivEarthS Labex program at Sorbonne Paris Cite (ANR-10-LABX-0023 and ANR-11-IDEX-0005-02).
\end{acknowledgements}
\bibliographystyle{aa}
\bibliography{biblio}

\begin{thebibliography}{45}
\expandafter\ifx\csname natexlab\endcsname\relax\def\natexlab#1{#1}\fi

\bibitem[{{Altamirano} \& {Belloni}(2012)}]{altamirano12}
{Altamirano}, D. \& {Belloni}, T. 2012, \apjl, 747, L4

\bibitem[{{Belloni} {et~al.}(2001){Belloni}, {M{\'e}ndez}, \&
  {S{\'a}nchez-Fern{\'a}ndez}}]{belloni2001}
{Belloni}, T., {M{\'e}ndez}, M., \& {S{\'a}nchez-Fern{\'a}ndez}, C. 2001, \aap,
  372, 551

\bibitem[{{Belloni} \& {Altamirano}(2013)}]{belloni_alt2013}
{Belloni}, T.~M. \& {Altamirano}, D. 2013, \mnras, 432, 19

\bibitem[{{Belloni} {et~al.}(2012){Belloni}, {Sanna}, \&
  {M{\'e}ndez}}]{belloni12}
{Belloni}, T.~M., {Sanna}, A., \& {M{\'e}ndez}, M. 2012, \mnras, 426, 1701

\bibitem[{{Casse} \& {Varniere}(2018)}]{CV18}
{Casse}, F. \& {Varniere}, P. 2018, mnras, 481, 2736

\bibitem[{{Casse} {et~al.}(2017){Casse}, {Varniere}, \& {Meliani}}]{CV17}
{Casse}, F., {Varniere}, P., \& {Meliani}, Z. 2017, \mnras, 464, 3704

\bibitem[{{Homan} {et~al.}(2003){Homan}, {Klein-Wolt}, {Rossi}, {Miller},
  {Wijnands}, {Belloni}, {van der Klis}, \& {Lewin}}]{Homan03}
{Homan}, J., {Klein-Wolt}, M., {Rossi}, S., {et~al.} 2003, \apj, 586, 1262

\bibitem[{{Kerr}(1963)}]{Kerr63}
{Kerr}, R.~P. 1963, Physical Review Letters, 11, 237

\bibitem[{{King} {et~al.}(2014){King}, {Walton}, {Miller}, {Barret}, {Boggs},
  {Christensen}, {Craig}, {Fabian}, {F{\"u}rst}, {Hailey}, {Harrison},
  {Krivonos}, {Mori}, {Natalucci}, {Stern}, {Tomsick}, \& {Zhang}}]{King14}
{King}, A.~L., {Walton}, D.~J., {Miller}, J.~M., {et~al.} 2014, \apjl, 784, L2

\bibitem[{{Klein-Wolt} {et~al.}(2004){Klein-Wolt}, {Homan}, \& {van der
  Klis}}]{KleinWolt04}
{Klein-Wolt}, M., {Homan}, J., \& {van der Klis}, M. 2004, Nuclear Physics B
  Proceedings Supplements, 132, 381

\bibitem[{{Kolehmainen} {et~al.}(2011){Kolehmainen}, {Done}, \& {D{\'\i}az
  Trigo}}]{KD11}
{Kolehmainen}, M., {Done}, C., \& {D{\'\i}az Trigo}, M. 2011, \mnras, 416, 311

\bibitem[{{Lin}(2012)}]{Lin12}
{Lin}, M.-K. 2012, \mnras, 426, 3211

\bibitem[{{Lovelace} \& {Hohlfeld}(1978)}]{Lovelace78}
{Lovelace}, R.~V.~E. \& {Hohlfeld}, R.~G. 1978, \apj, 221, 51

\bibitem[{{Lovelace} \& {Romanova}(2014)}]{Lovelace14}
{Lovelace}, R.~V.~E. \& {Romanova}, M.~M. 2014, Fluid Dynamics Research, 46,
  041401

\bibitem[{{Lyra} \& {Mac Low}(2012)}]{Lyra12}
{Lyra}, W. \& {Mac Low}, M.-M. 2012, \apj, 756, 62

\bibitem[{{McClintock} {et~al.}(2014){McClintock}, {Narayan}, \&
  {Steiner}}]{McC14}
{McClintock}, J.~E., {Narayan}, R., \& {Steiner}, J.~F. 2014, \ssr, 183, 295

\bibitem[{{Meheut} {et~al.}(2010){Meheut}, {Casse}, {Varniere}, \&
  {Tagger}}]{Meheut10}
{Meheut}, H., {Casse}, F., {Varniere}, P., \& {Tagger}, M. 2010, \aap, 516, A31

\bibitem[{{Meheut} {et~al.}(2012){Meheut}, {Keppens}, {Casse}, \&
  {Benz}}]{Meheut12}
{Meheut}, H., {Keppens}, R., {Casse}, F., \& {Benz}, W. 2012, \aap, 542, A9

\bibitem[{{Mignone} \& {McKinney}(2007)}]{Mignone07}
{Mignone}, A. \& {McKinney}, J.~C. 2007, \mnras, 378, 1118

\bibitem[{{Miller} {et~al.}(2002){Miller}, {Fabian}, {Wijnands}, {Reynolds},
  {Ehle}, {Freyberg}, {van der Klis}, {Lewin}, {Sanchez-Fernandez}, \&
  {Castro-Tirado}}]{Miller02}
{Miller}, J.~M., {Fabian}, A.~C., {Wijnands}, R., {et~al.} 2002, \apjl, 570,
  L69

\bibitem[{{Miller} {et~al.}(2001){Miller}, {Wijnands}, {Homan}, {Belloni},
  {Pooley}, {Corbel}, {Kouveliotou}, {van der Klis}, \&
  {Lewin}}]{miller01_1550}
{Miller}, J.~M., {Wijnands}, R., {Homan}, J., {et~al.} 2001, \apj, 563, 928

\bibitem[{Miniutti {et~al.}(2004)Miniutti, Fabian, \& Miller}]{Minutti04}
Miniutti, G., Fabian, A.~C., \& Miller, J.~M. 2004, Monthly Notices of the
  Royal Astronomical Society, 351, 466

\bibitem[{{Morningstar} \& {Miller}(2014)}]{MM14}
{Morningstar}, W.~R. \& {Miller}, J.~M. 2014, \apjl, 793, L33

\bibitem[{Reis {et~al.}(2009)Reis, Fabian, Ross, \& Miller}]{Reis09}
Reis, R.~C., Fabian, A.~C., Ross, R.~R., \& Miller, J.~M. 2009, Monthly Notices
  of the Royal Astronomical Society, 395, 1257

\bibitem[{{Reis} {et~al.}(2013){Reis}, {Miller}, {Reynolds}, {Fabian},
  {Walton}, {Cackett}, \& {Steiner}}]{Reis13}
{Reis}, R.~C., {Miller}, J.~M., {Reynolds}, M.~T., {et~al.} 2013, \apj, 763, 48

\bibitem[{{Remillard} \& {McClintock}(2006)}]{Remillard06}
{Remillard}, R.~A. \& {McClintock}, J.~E. 2006, \araa, 44, 49

\bibitem[{{Remillard} {et~al.}(2002){Remillard}, {Muno}, {McClintock}, \&
  {Orosz}}]{remillard02b}
{Remillard}, R.~A., {Muno}, M.~P., {McClintock}, J.~E., \& {Orosz}, J.~A. 2002,
  \apj, 580, 1030

\bibitem[{{Reynolds}(2014)}]{R14}
{Reynolds}, C.~S. 2014, \ssr, 183, 277

\bibitem[{{Shafee} {et~al.}(2006){Shafee}, {McClintock}, {Narayan}, {Davis},
  {Li}, \& {Remillard}}]{Shafee06}
{Shafee}, R., {McClintock}, J.~E., {Narayan}, R., {et~al.} 2006, \apjl, 636,
  L113

\bibitem[{{Sobczak} {et~al.}(2000){Sobczak}, {McClintock}, {Remillard}, {Cui},
  {Levine}, {Morgan}, {Orosz}, \& {Bailyn}}]{Sobczak00}
{Sobczak}, G.~J., {McClintock}, J.~E., {Remillard}, R.~A., {et~al.} 2000, \apj,
  544, 993

\bibitem[{{Steiner} {et~al.}(2012){Steiner}, {McClintock}, \&
  {Reid}}]{Steiner12}
{Steiner}, J.~F., {McClintock}, J.~E., \& {Reid}, M.~J. 2012, \apjl, 745, L7

\bibitem[{{Steiner} {et~al.}(2011){Steiner}, {Reis}, {McClintock}, {Narayan},
  {Remillard}, {Orosz}, {Gou}, {Fabian}, \& {Torres}}]{Steiner2011}
{Steiner}, J.~F., {Reis}, R.~C., {McClintock}, J.~E., {et~al.} 2011, \mnras,
  416, 941

\bibitem[{{Strohmayer}(2001)}]{S01a}
{Strohmayer}, T.~E. 2001, \apjl, 552, L49

\bibitem[{{Tagger} \& {Melia}(2006)}]{Tagger06}
{Tagger}, M. \& {Melia}, F. 2006, \apjl, 636, L33

\bibitem[{{Tagger} \& {Varni{\`e}re}(2006)}]{TV06}
{Tagger}, M. \& {Varni{\`e}re}, P. 2006, \apj, 652, 1457

\bibitem[{{Varniere} {et~al.}(2018){Varniere}, {Casse}, \&
  {Vincent}}]{VCH18XMM}
{Varniere}, P., {Casse}, F., \& {Vincent}, F.~H. 2018, in Proceedings of the
  XMM-Newton 2018 Science Workshop, TIME-DOMAIN ASTRONOMY: A HIGH ENERGY VIEW
  ESAC, MADRID, SPAIN, 13 - 15 JUNE 2018

\bibitem[{{Varniere} {et~al.}(2016){Varniere}, {Mignon-Risse}, \&
  {Rodriguez}}]{VMR16}
{Varniere}, P., {Mignon-Risse}, R., \& {Rodriguez}, J. 2016, \aap, 586, L6

\bibitem[{{Varniere} \& {Rodriguez}(2018)}]{VR18}
{Varniere}, P. \& {Rodriguez}, J. 2018, ApJ, 865, 113

\bibitem[{{Varni{\`e}re} \& {Tagger}(2006)}]{VT06}
{Varni{\`e}re}, P. \& {Tagger}, M. 2006, \aap, 446, L13

\bibitem[{{Varniere} {et~al.}(2011){Varniere}, {Tagger}, \&
  {Rodriguez}}]{VTR11}
{Varniere}, P., {Tagger}, M., \& {Rodriguez}, J. 2011, \aap, 525, A87

\bibitem[{{Varni{\`e}re} {et~al.}(2012){Varni{\`e}re}, {Tagger}, \&
  {Rodriguez}}]{VTR12}
{Varni{\`e}re}, P., {Tagger}, M., \& {Rodriguez}, J. 2012, \aap, 545, A40

\bibitem[{{Vincent} {et~al.}(2013){Vincent}, {Meheut}, {Varniere}, \&
  {Paumard}}]{Vin13}
{Vincent}, F.~H., {Meheut}, H., {Varniere}, P., \& {Paumard}, T. 2013, \aap,
  551, A54

\bibitem[{{Vincent} {et~al.}(2011){Vincent}, {Paumard}, {Gourgoulhon}, \&
  {Perrin}}]{Vin11}
{Vincent}, F.~H., {Paumard}, T., {Gourgoulhon}, E., \& {Perrin}, G. 2011,
  Classical and Quantum Gravity, 28, 225011

\bibitem[{{Vincent} {et~al.}(2014){Vincent}, {Paumard}, {Perrin}, {Varniere},
  {Casse}, {Eisenhauer}, {Gillessen}, \& {Armitage}}]{Vin14}
{Vincent}, F.~H., {Paumard}, T., {Perrin}, G., {et~al.} 2014, \mnras, 441, 3477

\bibitem[{{Walton} {et~al.}(2012){Walton}, {Reis}, {Cackett}, {Fabian}, \&
  {Miller}}]{Walton12}
{Walton}, D.~J., {Reis}, R.~C., {Cackett}, E.~M., {Fabian}, A.~C., \& {Miller},
  J.~M. 2012, \mnras, 422, 2510

\end{thebibliography}

\end{document}